\documentclass[conference]{IEEEtran}
\IEEEoverridecommandlockouts

\usepackage{cite}
\usepackage{amsmath,amssymb,amsfonts}
\usepackage{algorithmic}
\usepackage{graphicx}
\usepackage{textcomp}
\usepackage{xcolor}
\def\BibTeX{{\rm B\kern-.05em{\sc i\kern-.025em b}\kern-.08em
    T\kern-.1667em\lower.7ex\hbox{E}\kern-.125emX}}
\begin{document}

\title{Natural Language-Oriented Programming (NLOP):\\Towards Democratizing Software Creation
\thanks{Corresponding author: Prof. Amin Beheshti (amin.beheshti@mq.edu.au)\\
Accepted in: 2024 IEEE International Conference on Software Services Engineering (SSE), Shenzhen, China, July 7-13, 2024
}
}

\author{
\IEEEauthorblockN{Amin Beheshti*}
\IEEEauthorblockA{\textit{School of Computing} \\
\textit{Macquarie University}\\
Sydney, Australia \\
amin.beheshti@mq.edu.au}
}

\maketitle

\begin{abstract}

As generative Artificial Intelligence (AI) technologies evolve, they offer unprecedented potential to automate and enhance various tasks, including coding. Natural Language-Oriented Programming (NLOP), a vision introduced in this paper, harnesses this potential by allowing developers to articulate software requirements and logic in their natural language, thereby democratizing software creation.
This approach streamlines the development process and significantly lowers the barrier to entry for software engineering, making it feasible for non-experts to contribute effectively to software projects. By simplifying the transition from concept to code, NLOP can accelerate development cycles, enhance collaborative efforts, and reduce misunderstandings in requirement specifications.
This paper reviews various programming models, assesses their contributions and limitations, and highlights that natural language will be the new programming language.
Through this comparison, we illustrate how NLOP stands to transform the landscape of software engineering by fostering greater inclusivity and innovation.

\end{abstract}


\section{Introduction}

The field of software development has undergone numerous transformations, each aimed at addressing the specific challenges and limitations of the times. From the early days of ``Spaghetti Code'', characterized by its tangled and unstructured nature, to the more sophisticated models like Object-Oriented Programming (OOP) and Service-Oriented Computing (SOC), each shift in programming paradigm has sought to simplify complexities, enhance software maintainability, and improve developer productivity. This paper begins by tracing these evolutionary steps, providing a detailed review of the significant programming models that have shaped the modern software development landscape.

Building on the foundations of previous programming models, this paper introduces Natural Language-Oriented Programming (NLOP), a novel paradigm prompted by the advancements in Generative AI. NLOP utilizes natural language processing to allow programming with natural language, thus lowering barriers to entry and making software development accessible to a broader audience. By democratizing software creation, NLOP aims to expand the pool of software developers and encourages a more diverse range of problem solvers to participate in software innovation.

Language-Oriented Programming (LOP)~\cite{LOP0,LOP1,LOP2,LOP3} and Natural Language-Oriented Programming (NLOP) represent two different approaches to software development, each catering to distinct needs and audiences. LOP focuses on creating domain-specific languages (DSLs) that are meticulously designed to mirror the specifics of a given domain, requiring a deep understanding of language design principles and significant upfront investment. These languages aim to achieve a high degree of isomorphism with the domain, minimizing the gap between problem descriptions and implementations. On the other hand, NLOP leverages the universality and familiarity of natural human languages, utilizing advances in AI, particularly in natural language processing and generative AI, to interpret and convert spoken or written language into executable code. NLOP is designed to be more accessible, targeting a broader audience with no computer science background, thereby democratizing software development.

Natural Language-Oriented Programming (NLOP), introduced in this paper, offers a transformative approach to software development by integrating generative Artificial Intelligence (AI) with natural language processing to enable programming through natural language. Central to NLOP is its architecture, including a natural language interface, sophisticated prompt engineering for accurate AI interpretation, AI-driven code generation, and a microservices-based structure for enhanced modularity and scalability. This methodology not only democratizes software development by making it accessible to non-experts but also significantly reduces development time and costs. The business value of NLOP is profound, as it increases accessibility to software creation, enhances flexibility and scalability through its microservices architecture, and fosters improved collaboration across diverse teams. This paradigm shift holds the potential to revolutionize the software industry by making development more inclusive and aligned with business needs in the age of generative AI.

\begin{figure*}[t]
	\begin{center}
	\includegraphics[width=0.8\linewidth]{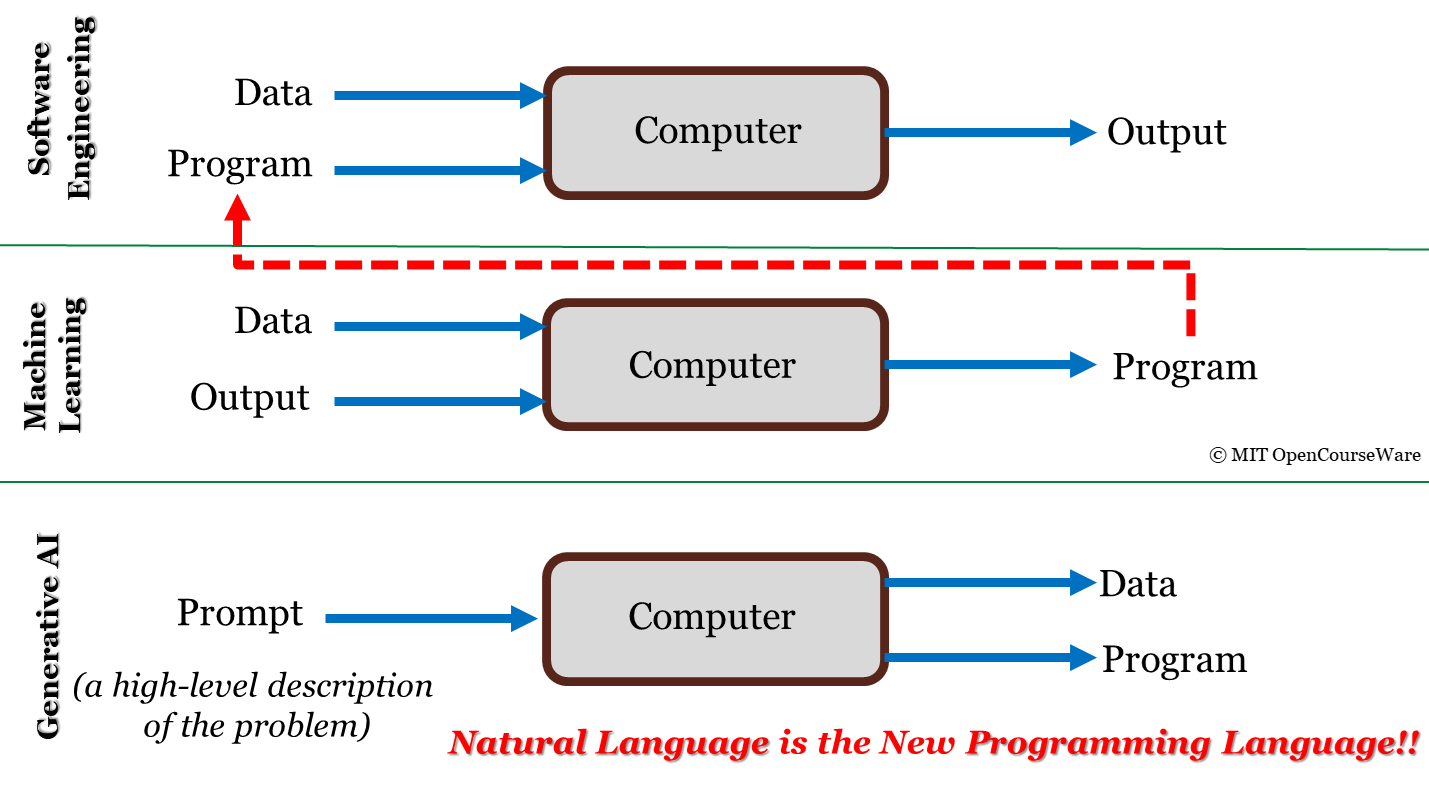}
	\end{center}
	\caption{Comparing Traditional Software Engineering, Machine Learning, and Natural Language-Oriented Programming Models}
	\label{traditionProgramming}
\end{figure*}

The rest of the paper is organized as follows:
In Sect.~II we compare Language-Oriented Programming (LOP) with
Natural Language-Oriented Programming (NLOP) model.
We provide an overview of programming in the age of Generative AI in Sect.~III.
Sect.~IV presents the evolution of programming paradigms.
In Sect.~V, we present the Natural Language-Oriented Programming (NLOP) framework and main components before concluding the paper in Sect.~VI.

\section{Language-Oriented Programming (LOP) vs. Natural Language-Oriented Programming (NLOP)}

Language-Oriented Programming (LOP)~\cite{LOP0,LOP1,LOP2,LOP3}, introduced to use constructed domain-specific languages designed specifically to address the problems and requirements of a particular domain. These languages are often new, constructed specifically for the project or problem at hand and require an understanding of language design principles. However, our proposed model, NLOP, employs existing natural languages and leverages the familiarity and universality of human language. This approach relies heavily on natural language processing (NLP) advances and generative AI to interpret and convert human language into executable code.

From the technical complexity and development focus, LOP involves significant upfront investment in language design and development, focusing on creating an optimal language that reflects the domain's specific processes as closely as possible. This might include defining syntax and semantics and possibly compiling or interpreting mechanisms tailored to the domain-specific language (DSL). However, NLOP focuses more on the interface between human language and the machine's ability to understand and execute it. It involves technologies like AI-driven prompt engineering, ProcessGPT~\cite{ProcessGPT}, AI code generation, and potentially sophisticated error handling and debugging mechanisms that interpret and manage ambiguously defined requirements.

From the Target audience point of view, LOP generally targets software developers and domain experts who are comfortable creating or learning new DSLs to solve domain-specific problems more effectively. However, NLOP model aims at a broader and more diverse audience, potentially including individuals with no traditional programming experience. It democratizes software development by allowing ideas to be expressed in natural language, thus lowering the entry barrier.

From an implementation and isomorphism point of view, LOP stresses creating a language as isomorphic as possible to the user’s domain, reducing the gap between the problem description and the implementation.
While NLOP also seeks to reduce the gap between intention and implementation, it does so through the flexible interpretation of natural language, aided by AI, rather than through the structured design of a new language.

\begin{figure*}[t]
	\begin{center}
	\includegraphics[width=0.9\linewidth]{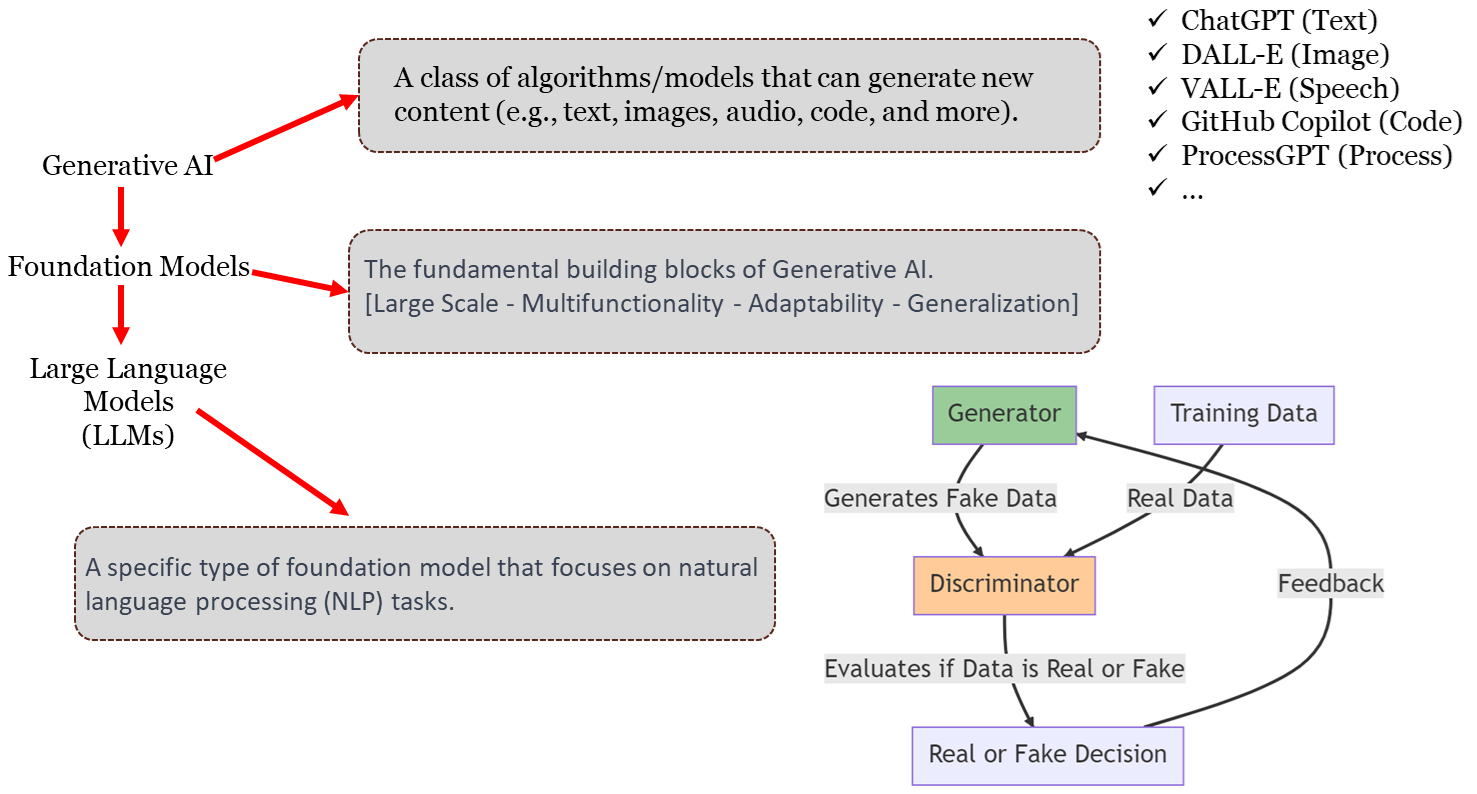}
	\end{center}
	\caption{Understanding Generative AI, Foundation Models, and Large Language Models (LLMs).}
	\label{GenAIFig}
\end{figure*}

\section{Programming: The Age of Generative AI}

Figure~\ref{traditionProgramming} compares traditional software engineering, Machine Learning, and programming in the age of generative AI Models.
In traditional software engineering, the development process is clear-cut: a predefined program written by developers processes the input data to produce the desired output. This model is grounded in meticulous planning, programming skills, and an understanding of algorithms. However, it is constrained by the requirement for explicit instructions for every possible scenario, often resulting in extensive and complex code bases that can be difficult to maintain and adapt.

Fast forward, Machine Learning (ML) models represent a paradigm shift by bypassing the need for explicit programming to perform specific tasks. Instead, an ML model is trained on a dataset and the corresponding outcomes, from which it learns to generalize and make predictions or decisions based on new data. This shift greatly reduces the manual effort in coding specific rules and allows the system to adapt to new data with minimal intervention. However, the success of this model hinges on the availability of large, high-quality datasets, and it typically requires specialized knowledge to fine-tune and interpret complex models.

Natural Language-Oriented Programming (NLOP), as conceptualized in this paper within the scope of Generative AI, takes a radical step forward. It entails providing a Generative AI component with a high-level description of the problem in natural language. The system then generates the desired data and/or the program that best fits the problem description. NLOP aims to enable individuals with a high-level understanding of their problems to create software solutions without deep technical expertise in coding or data science.

\subsection{Business Value}
The business values of NLOP are manifold. By lowering the barrier to entry for software creation, NLOP democratizes access to technology development, enabling a broader range of participants to innovate and solve problems through code generation and software creation. Organizations can leverage a wider talent pool, including those with domain expertise without traditional technical skills, to contribute directly to software development, fostering diversity in problem-solving and accelerating digital transformation. NLOP promises to dramatically reduce the time from concept to deployment, as the generative AI system can swiftly move from problem description to a working software solution.
NLOP's approach aligns closely with business agility requirements. It facilitates rapid prototyping and iteration (initiated by knowledge workers without a computer science background, ranging from end-users to business analysts and executive managers), allowing businesses to adapt quickly to changing market demands or user feedback. As generative AI continues to mature, NLOP is poised to streamline software production processes, enabling faster innovation cycles and potentially transforming the competitive landscape across industries.

\begin{figure*}[t]
	\begin{center}
	\includegraphics[width=1.0\linewidth]{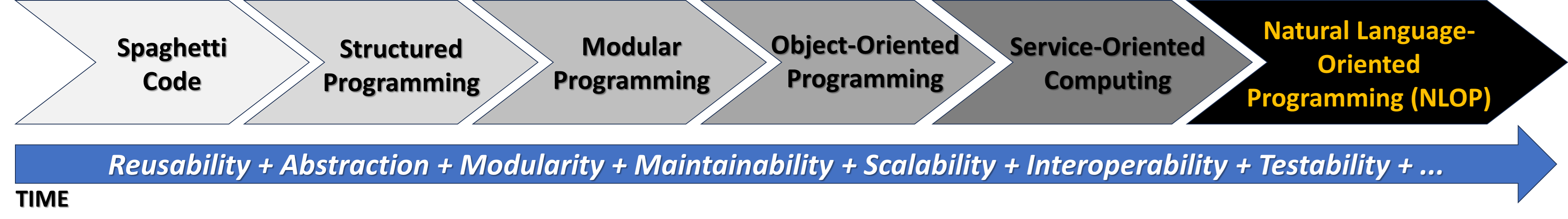}
	\end{center}
	\caption{Evolution of Programming Paradigms Over Time. This timeline illustrates the progression of major programming paradigms, highlighting key developments in language model features such as reusability and abstraction, which have significantly improved through the decades.}
	\label{timeFig}
\end{figure*}

\subsection{Generative Artificial Intelligence (AI)}

Generative AI encompasses a class of algorithms and models with the extraordinary capability to generate novel content~\cite{GenAI}. This includes various outputs such as text, images, audio, and executable code and processes. Examples of Generative AI at work include systems like ChatGPT\footnote{https://chat.openai.com/} for generating human-like text, DALL-E\footnote{https://openai.com/dall-e-2} for creating images from textual descriptions, GitHub Copilot\footnote{https://github.com/features/copilot}, which assists in coding by suggesting entire lines or blocks of code, ProcessGPT~\cite{ProcessGPT}, which generates new process models and instances when/if needed.
At the heart of Generative AI lies the Foundation Models~\cite{bommasani2021opportunities}. These sophisticated frameworks serve as the underlying architecture for building various generative systems. Foundation Models are trained on vast amounts of data and learn to capture this data's underlying patterns and structures. Once trained, they serve as a versatile starting point for numerous applications across different domains of AI, extending beyond just generative tasks.

Large Language Models (LLMs)~\cite{chang2024survey} are a subset of Foundation Models specialising in natural language processing tasks. LLMs, such as GPT (Generative Pre-trained Transformer~\cite{vaswani2017attention}) models, are designed to understand and generate human language. They can compose coherent and contextually relevant text often indistinguishable from text written by humans. Large Language Models are pivotal for language translation, question-answering, and summarization tasks.
Generative AI uses models that often follow a generator-discriminator architecture~\cite{creswell2018generative}. In this setup, the generator creates content intended to be indistinguishable from real data while the discriminator evaluates the generated content against actual data, providing feedback that helps the generator improve. Over time, through this feedback loop, the generator learns to produce results that can be very similar to authentic data, which is fundamental to the AI's ability to create believable and accurate content.

Figure~\ref{GenAIFig} presents the relationships between Generative AI, foundation models, and large language models (LLMs).
The impact of Generative AI and its related models on businesses and creative domains is profound. These technologies not only automate and enhance content creation but also enable personalized and engaging user experiences. From marketing and entertainment to software development and design, the potential applications of Generative AI are vast, opening up previously unimaginable possibilities. As these models continue to evolve, they promise to democratize creativity and innovation further, making them more accessible to everyone.

\section{Evolution of Programming Paradigms}

This section reviews various programming models, ranging from spaghetti code to object-oriented programming and service-oriented computing, evaluates their respective contributions and limitations, and underscores the emergence of natural language as the next evolution in programming languages.
Figure~\ref{timeFig} illustrates the evolution of programming paradigms over time.

\subsection{Spaghetti Code}

Spaghetti code, a term derived from its tangled, chaotic nature resembling a bowl of spaghetti, represents an early programming style characterized by several critical shortcomings. This style is notably marked by its lack of modular structure, leading to large blocks of code with minimal functional decomposition. Typically, spaghetti code relies heavily on \emph{goto} statements, which result in a non-linear flow of execution that is difficult to follow and predict. The absence of clear abstractions and encapsulations within this code style often results in high redundancy and repetition across the program. Furthermore, spaghetti code is characterized by high coupling and low cohesion among components, meaning that different parts of the program heavily depend on each other while not being focused on specific, well-defined tasks. These characteristics collectively contribute to significant difficulties in maintaining and extending spaghetti code, as understanding the flow and impact of changes can be challenging, increasing the risk of bugs and errors. As such, spaghetti code is widely regarded as problematic within the software development community due to its implications for software quality and maintainability.

\subsection{Structured Programming}

Structured programming emerged as a foundational approach to counteract spaghetti code's chaotic and tangled nature, offering a more logical and manageable framework for writing software. This paradigm promotes well-defined control structures such as assignments, selection statements (if and switch), and iteration statements (for, while, and do-while loops), which facilitate a linear and top-down development methodology. By advocating for these control structures, structured programming significantly enhances the readability and reliability of code.
While structured programming dramatically improved software development practices, it has limitations. One of the main criticisms of structured programming is its potential to lead to ``deep nesting," where multiple layers of loops and conditional statements are embedded within each other, complicating understanding and modification of the code. This can make the code verbose and difficult to maintain in complex scenarios. Additionally, structured programming sometimes requires more overhead in planning and implementing the control flow of complex programs, which can increase development time and effort.

\subsection{Modular Programming}

Modular programming is an advanced software development paradigm that builds upon structured programming principles to enhance code maintainability and readability further. It introduces the concept of modules—self-contained code units such as functions or procedures that encapsulate specific functionalities. This design philosophy encourages developers to decompose complex programs into discrete, manageable blocks that can be developed, tested, and debugged independently. Modular programming significantly reduces code redundancy and ensures consistency by promoting the reuse of modules across different parts of a program or even different programs. Localizing functionalities within modules limits the scope and impact of changes, simplifies maintenance tasks, and reduces the risk of introducing errors during updates or enhancements.
Despite its numerous benefits, modular programming does have its shortcomings. The division of a program into modules requires careful planning and a clear understanding of each module’s role and interface. This can introduce overhead in the design phase, requiring more upfront effort to define module boundaries precisely. Poorly designed module interfaces can lead to tight coupling between modules, undermining the benefits of encapsulation by making changes in one module propagate undesired effects to others. Furthermore, excessive modularization can lead to performance overheads due to increased inter-module communication or dependency resolution at runtime.

\subsection{Object-Oriented Programming}

Object-oriented programming (OOP) is a programming paradigm that extends the modular programming approach by introducing the concepts of classes and objects, thereby enhancing reusability and abstraction~\cite{wegner1990concepts}. In OOP, a class defines a blueprint for an object, encapsulating both data attributes and operations, i.e., functions that operate on the data. This structure enables a higher level of data abstraction by allowing programmers to create complex and abstract data types that behave as real-world objects. Classes support inheritance, allowing new classes to inherit attributes and methods from existing ones, promoting code reuse and reducing redundancy. Encapsulation shields data from direct access by external functions, enhancing modularity and preventing unintended interactions between different program parts.

Despite its advantages, OOP also has its limitations.
New programming models and enhancements to OOP have emerged to address these shortcomings. These include \textbf{aspect-oriented programming} (AOP), which aims to separate cross-cutting concerns from the main business logic, and \textbf{component-based programming}, which focuses on assembling pre-built, configurable, and reusable software components. Additionally, \textbf{functional programming} has gained popularity as an alternative that avoids mutable states and focuses on functions as the primary means of abstraction, offering benefits in scenarios requiring high concurrency and easy scalability. Each of these models attempts to refine the principles of OOP to overcome its challenges while leveraging its strengths in suitable contexts.

\subsection{Service-Oriented Computing}

Service-Oriented Computing (SOC)~\cite{papazoglou2003service} represents an evolutionary step in addressing some of the inherent shortcomings of earlier programming paradigms, including Object-Oriented Programming (OOP) and Aspect-Oriented Programming (AOP). SOC revolves around designing software in the form of services, i.e., well-defined self-contained modules that perform specific tasks and are designed to be independent of each other. These services communicate through standardized, network-based interfaces, typically realized as Web services or APIs. This approach fundamentally supports high reusability by allowing services to be used and repurposed across various applications without modification. SOC facilitates \emph{API engineering} and has laid the groundwork for \emph{Microservices Architecture}, wherein applications are built as collections of loosely coupled services, enhancing scalability and agility in development.

Service-Oriented Computing aims to simplify and standardize interactions between different software components, regardless of their underlying platforms or languages. This modularity allows developers to manage large systems more effectively by integrating small, manageable, and independently deployable services. SOC also addresses some limitations of AOP, such as the complexity of managing cross-cutting concerns across a large code base, by encapsulating these concerns into standalone services. While SOC brings significant benefits, it is not without challenges. The distributed nature of service-oriented architectures can introduce complexity in terms of deployment and management. Network latency, data integrity, and service availability issues are also critical. Furthermore, the granularity of services must be carefully managed to avoid excessive communications overhead or service bloat.

\textbf{Microservices Architecture} emerged as a strategic refinement of Service-Oriented Computing (SOC), specifically designed to address its shortcomings by emphasizing smaller, more granular services that operate independently yet communicate effectively through well-defined interfaces. This architecture enhances SOC by enabling better scalability, as each microservice can be deployed, scaled, and updated independently without affecting the overall system, thereby reducing downtime and improving performance. A microservices architecture also facilitates agile development and continuous delivery by allowing teams to develop and deploy services independently, thus speeding up iteration cycles and reducing the coordination overhead required in monolithic SOC applications. Additionally, by allowing each microservice to use its own technology stack and data storage solutions, the architecture alleviates the constraints of a uniform development environment and centralized data management, thus enhancing flexibility and resilience. These improvements address critical pain points in traditional SOC by decentralizing functionalities and data, which minimizes single points of failure and optimizes resource utilization across distributed systems.

\section{Natural Language-Oriented Programming (NLOP): a New Computer Programming Model}

\begin{figure*}[t]
	\begin{center}
	\includegraphics[width=1.05\linewidth]{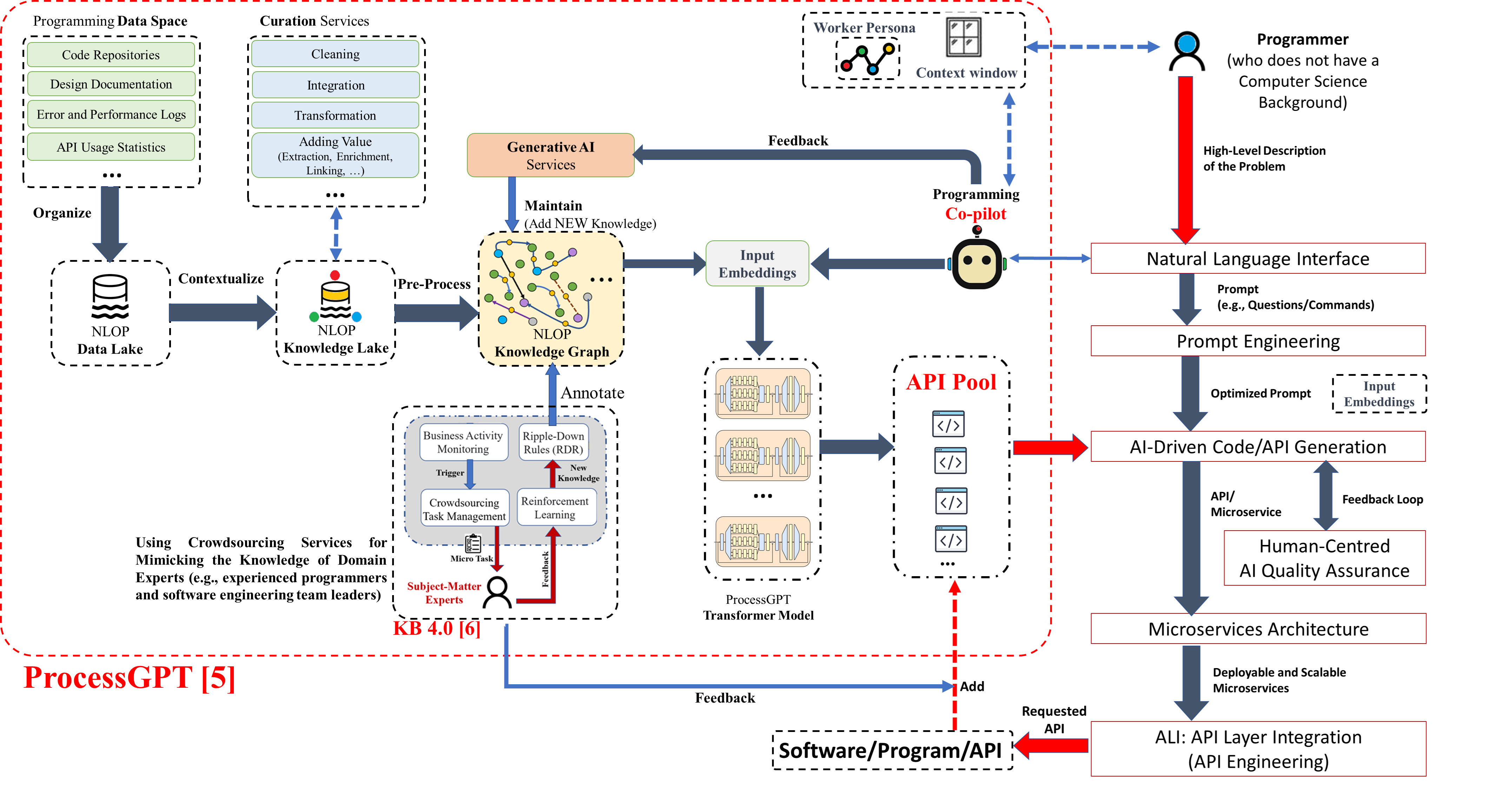}
	\end{center}
	\caption{Proposed framework for Natural Language-Oriented Programming.}
	\label{NLOP}
\end{figure*}

Natural Language-Oriented Programming (NLOP) is a visionary software development paradigm that harnesses the power of generative artificial intelligence (AI) to interpret and convert natural language into executable code.
The goal is to make software development more accessible to a broader range of people, irrespective of their technical skills or programming expertise.
The NLOP methodology involves several key components forming a new software development architecture, primarily designed to make programming more accessible and efficient.
Figure~\ref{NLOP} illustrates key components, each contributing to a unique architecture designed to streamline software development. In the following, we discuss each component in more detail.

\subsection{ProcessGPT: Understanding Programming Data Spaces for NLOP}

ProcessGPT~\cite{ProcessGPT}, our previous work. a novel application of generative pre-trained transformers is adept at transforming extensive and diverse programming data spaces (i.e., including Code Repositories, Design Documentation, Error and Performance Logs, API Usage Statistics, User Interaction Data, and Deployment and Operations Data) into finely contextualized and reusable computer programs, represented as APIs and Microservices. This transformation is pivotal for fueling the backbone of our NLOP model.

ProcessGPT ingests and processes vast arrays of programming data to generate a comprehensive pool of APIs. This pool forms the core architecture upon which the NLOP system is built. By leveraging deep learning techniques and a transformer-based architecture, ProcessGPT effectively captures and translates the intricacies and semantics of raw programming data into highly functional and scalable APIs. These APIs are then catalogued and accessible for on-the-fly usage in programming and software development workflows.

\subsection{NLI: Natural Language Interface}

This component is essential to the NLOP framework, designed to bridge the gap between human linguistic expression and machine code generation.
%
The primary function of the NLI is to accurately interpret and translate natural language inputs into structured programming commands. It leverages advanced natural language processing (NLP) techniques and co-pilot technology~\cite{pudari2023copilot} to understand user commands' context, intent, and specifics, ensuring that the AI-generated code aligns closely with the user’s requirements. The key features and future research directions may include:

\begin{itemize}
    \item  Advanced NLP Capabilities: Utilizing state-of-the-art NLP models, the NLI can comprehend a wide range of natural language expressions, including complex instructions and technical descriptions. This allows for a more intuitive and accessible programming process that accommodates various user expertise levels.
    \item Contextual Understanding: The NLI is equipped with contextual understanding capabilities to grasp the broader context of a software project, enabling it to make informed decisions about code generation and to handle ambiguous or incomplete instructions more effectively.
    \item Error Handling and Suggestions: The NLI includes robust error-handling mechanisms that identify potential misunderstandings or errors in natural language inputs to enhance usability. It offers corrective suggestions and seeks clarifications to ensure the accuracy of the information being processed.
    \item Integration with Development Environments: The interface is designed to integrate seamlessly with existing development environments and tools, facilitating a smooth transition from natural language input to code generation and deployment.
    \item User Feedback Loop (RLHF+RLAIF~\cite{lee2023rlaif}): An integral feature of the NLI is its user feedback loop that focuses on scaling reinforcement learning from human feedback with AI feedback~\cite{lee2023rlaif}, for continuously improving the accuracy and effectiveness of the NLP models.
\end{itemize}

The Natural Language Interface significantly lowers the technical barrier to software development, making it possible for individuals without traditional programming skills to participate actively in creating and managing software projects. By democratizing access to programming, the NLI expands the talent pool and fosters a more inclusive and innovative development environment. 

\subsection{PE: Prompt Engineering}

This component enhances the precision with which the AI interprets and acts on user instructions, ensuring that the software developed closely aligns with the intended functionality.
Prompt Engineering involves carefully crafting and optimising user prompts to guide the AI's understanding and response. This process is crucial for translating user inputs, which can be vague or contextually complex, into clear, actionable programming tasks that the AI can execute accurately.
The key features and future research directions may include:

\begin{itemize}
    \item Optimization of User Inputs: This feature refines natural language inputs into well-defined prompts structured to elicit the most accurate and relevant responses from the AI. This involves parsing user inputs, extracting key information, and reformulating them into prompts optimized for AI comprehension.
    \item Context Management: This component incorporates techniques to manage and utilize contextual information from the ongoing project or previous interactions. This enables the AI to maintain continuity and relevance in its code generation, adapting its outputs based on the broader context of the software development project.
    \item Dynamic Adaptation: Prompt Engineering dynamically adapts the prompts based on real-time interactions and user corrections. This flexibility ensures that the AI remains aligned with the user’s evolving requirements throughout the software development process.  An integral aspect of prompt engineering is incorporating feedback from previous AI outputs (RLHF+RLAIF~\cite{lee2023rlaif}).
    \item Quality Assurance: This feature routinely checks the effectiveness of different prompts and their impact on the AI's performance. The component can identify the most effective strategies for communicating with the AI by analysing the success rates of various prompt formulations.
\end{itemize}

Prompt Engineering significantly enhances the reliability and efficiency of AI-driven software development. By ensuring that AI systems accurately understand and execute user instructions, this component reduces the potential for errors and misinterpretations, leading to faster development cycles and higher-quality software. 

\subsection{AI-CG: AI-Driven Code Generation}

This component is the central element of the NLOP framework. AI-CG transforms natural language inputs into executable software code that will be encapsulated as an API. This component leverages AI technologies to automate coding~\cite{pudari2023copilot}, and will be fed by the ProcessGPT component (Figure~\ref{NLOP}) to bridge the gap between conceptual design and functional software.
AI-driven code Generation primarily functions by interpreting optimized prompts from the Prompt Engineering component and generating corresponding code snippets or entire modules. It utilizes machine learning algorithms and models trained on vast code datasets to produce accurate, efficient, and maintainable code that adheres to industry standards.
The key features and future research directions may include:

\begin{itemize}
    \item Advanced Machine Learning Models: This component employs neural networks~\cite{aggarwal2018neural} and other AI models trained on diverse programming languages and frameworks. These models enable the system to understand complex programming tasks and generate appropriate code based on natural language descriptions.
    \item Multi-language Support: AI-driven code Generation supports multiple programming languages, allowing users to specify or switch the output language as per project requirements. This flexibility ensures the component can cater to various software development environments and applications.
    \item Integration with Development Tools: The code generation engine is designed to integrate seamlessly with development tools and IDEs (Integrated Development Environments). This integration facilitates immediate testing, deployment, and iteration of the generated code within the existing workflows of software development teams.
    \item Quality and Security Assurance: Embedded within the AI code generator are mechanisms for ensuring that the code works effectively and is secure and optimized. It includes static code analysis tools to detect vulnerabilities and inefficiencies before the code is deployed.
    \item Customization and Scalability: The component allows for customization of the code generation process based on user preferences and project-specific guidelines. It can scale its operations to handle large-scale projects involving extensive codebases and complex functionalities.
\end{itemize}

The AI-Driven Code Generation component significantly accelerates software development by automating one of its most labour-intensive aspects: writing and testing code. By reducing the time and resources required for coding, businesses can achieve faster time-to-market for their software products. Additionally, the high quality and reliability of the AI-generated code help minimize post-deployment issues, enhancing product stability and user satisfaction. This component streamlines software development and enables companies to leverage their human resources more effectively, allocating more time to creative and strategic tasks rather than routine coding.

\subsection{HCAI-QA: Human-Centred AI Quality Assurance}

This component uses Reinforcement Learning with Human/AI Feedback (RLHF/RLAIF~\cite{lee2023rlaif}) to: (i)~ensure the integrity and effectiveness of AI-driven code generation; and (ii)~merge human expertise with automated processes to oversee and enhance the quality of software modules produced by artificial intelligence. HCAI-QA operates by continuously monitoring the output of AI code generators, assessing the code for adherence to functional and performance standards. HCAI-QA employs a combination of real-time error checking, performance evaluation, and compliance with coding standards to ensure that the generated code is functional but also optimized and maintainable. The key features and future research directions may include:

\begin{itemize}
    \item Interactive Feedback Loop: We leverage our previous work, KB~4.0~\cite{GenAI}, to directly incorporate Subject-Matter Expert's (SME) feedback into the AI's learning process. This feedback is gathered through systematic reviews of the AI-generated code by developers, who provide corrective insights to refine and train the AI models under Reinforcement Learning from Human Feedback (RLHF) protocols~\cite{lee2023rlaif}.
    \item Automated Testing Integration: HCAI-QA integrates comprehensive automated testing tools that execute a series of tests on the AI-generated code, including unit, integration, and system tests, to ensure robustness and functionality before deployment.
    \item Performance Metrics Tracking: The component tracks performance metrics such as execution efficiency, resource consumption, and compliance with best practices, providing insights that help continually optimise the AI algorithms.
    \item Version Control and Anomaly Detection: Integration with version control systems allows for tracking iterations and modifications effectively, facilitating easier rollbacks and historical comparisons. Additionally, advanced anomaly detection techniques are utilized to identify and address unusual patterns in the code that may signify underlying issues.
\end{itemize}

By embedding human and/or artificial intelligence into the loop, HCAI-QA elevates the reliability of AI-generated code and aligns the development process more closely with human-centric design principles. This ensures the software developed is user-focused, secure, and highly maintainable, reducing downstream costs and enhancing the overall development lifecycle.

\subsection{MA: Microservices Architecture}

This component is fundamental to the NLOP framework, designed to structure the AI-generated code into small, independently deployable services. This architectural style facilitates scalable, flexible, and robust software systems.
Microservices Architecture breaks down the software application into a collection of smaller services that run independently and communicate with each other via well-defined APIs. Each microservice is self-contained and responsible for performing a unique business function. This decomposition aligns with modern software engineering practices, which favour agility and efficiency in development, deployment, and maintenance.
The Microservices Architecture component is pivotal in structuring and managing the software system. The microservices within this architecture are dynamically generated by the ``AI-Driven Code Generation" component, which efficiently translates natural language inputs into standalone services. Each service is crafted to perform specific business functions independently, allowing for a modular and scalable system design. 

The Microservices Architecture component is tasked with comprehensive API engineering for these services. It manages a large pool of APIs, ensuring they are well-defined, properly documented (auto-generated), and maintainable. This management includes handling API dependencies, versioning, and ensuring seamless service communication. Doing so facilitates robust integration and interaction among the microservices, enabling them to function cohesively as part of a larger system. 
The key features and future research directions may include:

\begin{itemize}
    \item Service Independence: Each microservice is developed, deployed, and maintained independently. This independence allows for individual scaling, updating, or bug fixing of services without impacting the entire system, enhancing system resilience and flexibility.
    \item Technology Diversity: Microservices can be written in different programming languages and use different data storage technologies based on what is best suited for their specific requirements. This technological heterogeneity allows teams to utilize the best tools for specific tasks, optimizing performance and functionality.
    \item Decentralized Management: Instead of relying on a single system, each microservice is decentralized, which helps isolate and contain data-related issues to the respective service. This approach enhances the security and stability of the overall system.
    \item Automated Deployment: Microservices support automated, continuous integration and deployment practices. This capability is crucial for maintaining high velocity in development cycles and enabling quick releases with minimal downtime.
    \item API Engineering and Version Management: Effective API engineering includes version management, which is crucial for maintaining multiple versions of APIs to support legacy systems while enabling incremental improvements and updates without disrupting existing clients. This strategic approach to API lifecycle management allows businesses to evolve their services dynamically and maintain compatibility across different system versions, thereby reducing client-side friction and enhancing overall service adaptability.
\end{itemize}

Microservices Architecture significantly contributes to the agility and scalability of business operations. By allowing businesses to develop and deploy parts of their application independently, they can respond more rapidly to market changes, customer demands, or emerging technologies. Additionally, this architecture reduces the complexity of large monolithic systems, making it easier to manage and evolve. The modular nature of microservices also fosters a more collaborative and productive development environment, as teams can focus on specific areas of expertise without burdening the entire system's intricacies. This leads to faster development cycles, increased productivity, and a more dynamic and innovative approach to software development.

\subsection{ALI: API Layer Integration}

This component is an essential feature of the Natural Language-Oriented Programming (NLOP) framework, specifically designed to facilitate communication and data exchange between the independently functioning microservices and the external world. This component is pivotal in ensuring that the microservices architecture operates cohesively and efficiently.
API Layer Integration acts as the communicative glue that binds together different microservices by providing standardized interfaces for data interchange. It allows microservices to expose their functionalities to other internal services or external clients without exposing the underlying complexities. This abstraction layer helps maintain loose coupling between services, which is fundamental to achieving the scalability, flexibility, and resilience.
The key features and future research directions may include:

\begin{itemize}
    \item Standardization of Communication: The API Layer provides a uniform way to access the diverse functionalities of individual microservices. Standardizing how services communicate simplifies integration and interaction among services.
    \item Security Enforcement: This component includes robust security protocols such as authentication, authorization, and encryption to ensure secure interactions between services and external clients. This is crucial for protecting data integrity and privacy within the microservices ecosystem.
    \item Load Balancing and Fault Tolerance: API Layer Integration supports mechanisms like load balancing and fault tolerance to distribute incoming requests efficiently across services and to manage failures gracefully. This ensures high availability and reliability of the services even under high load or during partial system failures.
    \item Service Monitoring: It facilitates monitoring service health, usage, and performance metrics, allowing for proactive management of the system's operational state.
    \item API Gateway: This gateway provides a single, unified API entry point across numerous microservices. It handles request routing, composition, and protocol translation, enabling clients to interact seamlessly with various microservices without managing multiple endpoints. Additionally, the API Gateway offers important functionalities such as rate limiting, authentication, and authorization, which are essential for securing and managing access to the microservices. This layer simplifies the complexity of interacting with a distributed system, ensuring that the architecture remains scalable and manageable despite the increasing number of services.
\end{itemize}

The quality of the APIs generated at this phase will be reviewed by the subject-matter expert's feedback loop and the final product will be added to the API pool component in ProcessGPT (Figure~\ref{NLOP}).

\subsection{Enhancing API Layer Integration with ProcessGPT}

When integrated with the API Layer, ProcessGPT serves a dual role: as a dynamic generator of API specifications based on the evolving needs of microservices and as an intelligent decision-maker, assisting in the routing, composing, and translating protocol interactions between services.
The coupling of ProcessGPT with API Layer Integration translates to an environment where microservices are not only automatically generated but are also endowed with the capability to expose their functionalities through well-constructed APIs. ProcessGPT can help standardise these communication interfaces, ensuring they are robust, secure, and conducive to high levels of interoperability and developer experience. As a result, the system becomes adept at simplifying the complexities inherent in microservices communication, reinforcing the architectural cohesion.

Security, a non-negotiable aspect of API Layer Integration, is bolstered through ProcessGPT's predictive models. These models preemptively identify and address potential vulnerabilities, reinforcing authentication, authorization, and encryption measures. Moreover, by tapping into the wealth of process knowledge, ProcessGPT can intelligently manage API traffic, implementing rate limiting and fault tolerance to maintain service availability and resilience.
ProcessGPT's most notable contribution could be in API Gateway management. It provides suggestions for efficient request routing and facilitates the aggregation of services, acting as an intelligent intermediary that understands the context and the required composite functionalities. ProcessGPT's ability to learn and adapt makes it an invaluable asset in overseeing the seamless execution of complex service orchestrations.
ProcessGPT has the potential to revolutionize the API ecosystem within NLOP, making it more intuitive, responsive, and tailored to businesses' real-time needs.


\section{Conclusion and Future Work}

The introduction of Natural Language-Oriented Programming (NLOP) marks a significant advancement in software development. NLOP embraces the potential of generative AI to transform coding into an inclusive, accessible task. By integrating natural language processing capabilities, NLOP allows individuals with no programming backgrounds to contribute to and innovate within the software development process. This paradigm shift democratises software creation and enhances the agility and efficiency of development projects by streamlining the translation of conceptual ideas into executable code.
Looking forward, several areas of NLOP require further exploration and development:

\textbf{From ProcessGPT to NLOP:}
ProcessGPT~\cite{ProcessGPT} emerges as a revolutionary force in the Natural Language-Oriented Programming (NLOP) domain, specifically within the Microservices Architecture component. ProcessGPT's integration within the NLOP framework signifies a leap towards automated, dynamic generation and maintenance of microservices based on contextualized understanding and sophisticated decision-making processes informed by substantial datasets and expert knowledge.
Within the Microservices Architecture component of NLOP, ProcessGPT is a generative engine that constructs discrete, modular services, each designed to perform distinct business functions autonomously. These microservices are not only generated based on a high-level description of problems. Still, they are also iteratively refined through continuous learning, tapping into ProcessGPT's proficiency in understanding and processing complex workflows and business logic.

The design of ProcessGPT facilitates a granular and decentralized data management strategy, which is critical in the Microservices Architecture. By fostering service independence, ProcessGPT enables individuals to scale and update services without causing systemic disruptions.
%
As businesses navigate the complexities of digital transformation, ProcessGPT empowers collaborative environments and leads to more rapid and productive development cycles.

\textbf{Integration with Augmented and Virtual Reality (AR/VR):}
NLOP could be expanded to incorporate AR/VR platforms, blending natural language with user imagination to formulate prompts precisely.
This will enable: Enhanced Cognitive Understanding, Collaborative Coding, and Rapid Prototyping.

\textbf{Quantum Computing Integration:} As quantum computing advances, there is a potential for NLOP to be extended to quantum software development.

\textbf{Ethical AI Code Generation:} As AI systems take on more coding tasks, embedding ethical considerations into code generation becomes crucial. Future NLOP systems could be equipped with ethical guidelines to ensure that the code they generate adheres to ethical standards and practices, preventing biases and promoting fairness and transparency.

\textbf{Neuro-Symbolic AI for Code Synthesis:} Integrating neuro-symbolic AI approaches could enhance NLOP systems' ability to understand and implement complex problem-solving strategies expressed in natural language. This hybrid approach combines the neural network's learning capabilities with symbolic AI's reasoning, improving the system's robustness and effectiveness.

\textbf{Collaborative NLOP Environments:} Developing collaborative platforms where both AI and human developers can interact dynamically to build and refine code.

\textbf{Personalized AI Coding Assistants:} NLOP systems could evolve into personalized AI assistants that understand individual developer preferences and styles and adapt their code generation accordingly.

\textbf{Adaptive Learning Systems for Code Optimization:} Future NLOP systems could incorporate adaptive learning algorithms that generate code and refine it over time based on performance metrics and user feedback. These systems would learn from each interaction, continuously improving the efficiency, readability, and functionality of the code.

\bibliographystyle{IEEEtran}
\bibliography{bibliography}

\begin{IEEEbiography}[{\includegraphics[width=1in,height=1.25in,clip,keepaspectratio]{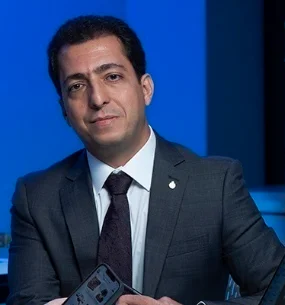}}]{Amin Beheshti}
{\space}holds a B.S. (1st Hons.) and M.S. degrees (1st Hons.) in computer science and engineering and a Ph.D. in computer science from UNSW Sydney, Australia. Amin is a Full Professor of data science at Macquarie University. He is currently the Director of the Centre for Applied Artificial Intelligence and the Head of the Data Science Research Laboratory, School of Computing, Macquarie University. He is a leading author of several books in data, social, and process analytics, co-authored with other high-profile researchers.
\end{IEEEbiography}

\end{document}